# Increasing Spatial Fidelity and SNR of 4D-STEM Using Multi-Frame Data Fusion

*Colum M. O'Leary[†,‡,1], Benedikt Haas[†,2], Christoph T. Koch[2], Peter D. Nellist[1], Lewys Jones*[3,4],*

1: Department of Materials, University of Oxford, Oxford, UK
2: Institut für Physik & IRIS Adlershof, Humboldt-Universität zu Berlin, Berlin, Germany
3: Advanced Microscopy Laboratory, Centre for Research on Adaptive Nanostructures & Nanodevices (CRANN), Dublin, Ireland
4: School of Physics, Trinity College Dublin, Dublin, Ireland

*4D-STEM, in which the 2D diffraction plane is captured for each 2D scan position in the scanning transmission electron microscope (STEM) using a pixelated detector, is complementing, and increasingly replacing existing imaging approaches. However, at present the speed of those detectors, although having drastically improved in the recent years, is still 100 to 1,000 times slower than the current PMT technology operators are used to. Regrettably, this means environmental scanning-distortion often limits the overall performance of the recorded 4D data. Here we present an extension of existing STEM distortion correction techniques for the treatment of 4D-data series. Although applicable to 4D-data in general, we use electron ptychography and electric-field mapping as model cases and demonstrate an improvement in spatial-fidelity, signal-to-noise ratio (SNR), phase-precision and spatial-resolution.*

† these authors contributed equally

‡ now at University of California, Los Angeles

* corresponding author: lewys.jones@tcd.ie

## Introduction

The scanning transmission electron microscope (STEM) is an exceptionally powerful instrument for materials characterization. By forming an electron probe at the sample, a plethora of spatially localized transmitted, reflected, or emitted imaging and spectroscopic signals are available simultaneously to the operator. Of these signals annular dark-field, bright-field, or annular bright-field (ADF, BF, ABF respectively) are some of the most popular imaging techniques. Each of these modalities are commonly realized using a scintillator-photomultiplier style detector which collects a particular scattering angle range within the detector plane. The electron-flux falling on the scintillator leads to a voltage at the photomultiplier (PMT) which is expressed as a function of position to yield an image. However, more information per scan point can be captured when using a pixelated detector instead of an integrating detector. While early implementation of this '4D-STEM' – in which reciprocal space is sampled in 2D while real space is scanned in 2D – happened already in the early 2000s [1,2], the slow speed of the cameras at that time limited the spread of this technique or the real-space field-of-view [3]. Recent advances in fast electron sensors now allow for pixelated detectors to acquire data 10–100 times faster than slow scan charge-coupled devices. A trivial use of these new sensors would be to synthesize the existing imaging modes previously available; however, the abundance of information available in a 4D-STEM data set has enabled a variety of imaging and reconstruction techniques to be practically realized. These applications include mapping crystalline orientation [4–7], mapping electric fields [8–10] and magnetic domains [11,12], using center-of-mass (COM) approaches [8,13], and performing phase reconstruction techniques such as electron ptychography [14–16]. More past and current applications of 4D-STEM and related hardware can be found in reviews by Ophus [17] and MacLaren [18].

In recent years, both ptychography and COM-imaging have become of especial interest for imaging of light and thin specimens where absorption contrast may be very weak at up to atomic-resolution. The object phase may carry information about the sample potential, electric field and atomic resolution polarization. Focused-probe STEM ptychography approaches have already demonstrated a variety of novel applications, such as visualizing lithium in cathode materials [19], imaging beam-sensitive zeolites at atomic resolution [20], optical-sectioning of carbon-based nanostructures [21], and increasing the resolution beyond that of conventional STEM images [22], even when severely binning the data for achieving high scan speeds [23]. COM analysis of atomic resolution 4D-STEM data has been utilized in equally diverse applications, including the identification of surface adatoms [24] and the determination of charge accumulation at interfaces [10].

4D-STEM detectors need not wholly replace conventional hardware and can be installed co-operatively to record the beam which passes through the center of an annular detector. This configuration, along with the additional flexibility of variable camera-lengths, allows for the simultaneous acquisition of multiple detector signals. For example, by acquiring ADF and ptychographic data simultaneously, both light and heavy elements can be identified using data from a single experimental STEM scan [21]. Furthermore, spectroscopic data can be acquired simultaneously to 4D-STEM and annular detector imaging by using an off-axial detector.

High-resolution TEM focal/tilt series restoration (HRTEM-FTSR) offers one route to fusing contrast transfer functions and achieving phase-imaging [25,26]. However, it requires the precise lateral registration of defocused frames and contrast reversals mean this remains a challenging task. With 4D-STEM approaches, all beam-angles (tilts) are recorded simultaneously, and a phase image can be reconstructed from a single scan-frame. However as discussed, the trade-off is the risk of line-by-line scanning instability which HRTEM is not affected by in the same way.

In the STEM, unlike a conventional transmission electron microscope (CTEM), a raster scan is required to build up a 2D image or chemical map. The sequential nature of this serial-scan opens the STEM to the potential weakness of imperfect environmental stability during the time needed to record each scan-frame. These instabilities might be acoustic, seismic, thermal, barometric, electronic or magnetic in nature and should be kept to an absolute minimum wherever possible through good instrument and suite design [27,28]. In this manuscript, although the approach followed is general, we will consider the case of atomic-resolution STEM as the high magnification presents the worst-case scenario for scan-distortions.

In the data, these artefacts manifest themselves in a variety of ways; stage movements *between* frame-captures appear as a simple rigid-translation, whereas continual stage-drift *within* a frame's recording time (but still slow with respect to the frame time, e.g. <¼ period per frame-time) will add an affine shearing to the image data [29–31]. Slightly faster stage movements or environmental distortions, with periods ranging from around 1 cycle-per-frame up to several tens of cycles-per-frame, will appear as an irregular non-linear warping whose shifts are characteristically highly correlated along the fast-scan direction [32–35]. Higher-frequency instabilities can also arise in the microscope lab, but these can be relatively easily damped with the use of instrument enclosures or heavy curtains.

Unfortunately, even with the newest generation of fast pixelated STEM sensors, the pixel-times (and hence frame rates) for 4D-STEM may be 100-1000 times slower than simple integrating type detectors [10]. This results in the distortion frequencies of interest being slower by the same ratio, more challenging to shield or damp in hardware, and almost unavoidable in experimental data.

In this manuscript, we propose a multi-frame acquisition and data registration strategy to compensate for the effects of scanning distortions in 4D-STEM. By acquiring multiple frames and subsequently aligning them, both the spatial precision and signal-to-noise ratio (SNR) of the data can be improved. This, in turn, improves the quality of the reconstructions obtained from 4D-STEM imaging and analysis techniques such as electron ptychography and center-of-mass imaging.

The remainder of this manuscript is structured as follows; firstly, the frequency ranges of scanning-distortion most deleterious to atomic resolution 4D-STEM mapping will be introduced. This is followed by a discussion of the data acquisition and non-rigid correction techniques used to mitigate these scanning-distortions. Next, we present two example applications of the proposed workflow: multi-frame electron ptychography and COM imaging of monolayer, bilayer and bulk-like crystalline materials.

Following this, a discussion of the attained spatial-resolution and phase/field precision will be presented. Finally, the potential applications of this correction approach will be discussed.

## Background

In addition to the imaging and diffraction capabilities of the STEM, various spectroscopic modes are also available, such as energy-dispersive x-ray spectroscopy (EDX) and electron energy-loss spectroscopy (EELS). Although annular image detectors, EDX spectrometers and EELS systems can be used to acquire data simultaneously, the optimal acquisition times for each technique are not identical.

Typical frame times of conventional STEM imaging range from one second to a few tens of seconds. As a result, these scans can be vulnerable to high-frequency "scan-noise" (Figure 1, ~100Hz up to few kHz) and software corrections for this have been proposed [30,36,37]. Alongside these software tools, improved efforts in microscope rooms (such as heavy curtains, vibration damping systems, electromagnetic shielding/compensation, etc.) and from instrument manufacturers to better block barometric fluctuations and acoustic noise (such as the goniometer 'clamshell' or full enclosures) have largely solved this issue.

However, for spectroscopic acquisitions, either because of hardware limitations or simply to collect enough signal, pixel times and hence frame times are often 1-2 orders of magnitude larger than for conventional STEM imaging. As a result, the frequencies that instruments are susceptible to are shifted lower and many operators will be familiar with this when recording spectral maps. Typical frequency ranges for imaging and spectroscopy are shown in Figure 1.

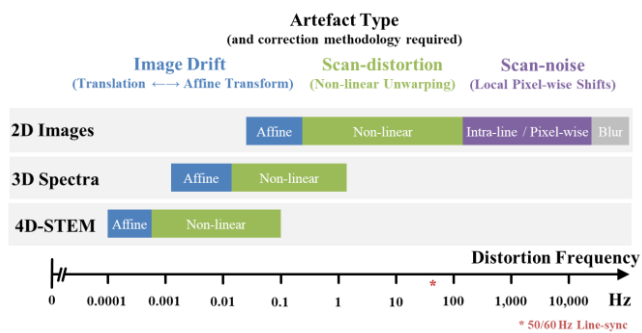

*Figure 1. Comparison of the frequency ranges of various scan-induced data artefacts and the associated correction methodologies. For the slower recording strategies, the same manifestations are shifted down to lower frequencies.*

For fully pixelated STEM sensors, the read-out rate is again another order of magnitude or more slower. Typical readouts may be only 1k -8k frames per second (where now one frame refers to one STEM probe position). Thus, electron ptychography may be susceptible to environmental and instrumental distortions as slow as 1mHz. Even with the advent of faster detector technology, the limited dynamic range of many fully pixelated STEM sensors places restrictions on the maximum SNR obtainable for a single data set. Thus, there is currently a need for data registration of multi-frame 4D-STEM data.

Fortunately, the correction of scanning distortion has received significant study which can be repurposed for this new field. The scanned physical-probe community (including atomic-force microscopy [32], and scanning-tunneling microscopy [29,38]) have each developed techniques for remedying imperfect scanning. More recently the electron imaging community has developed similar tools to push the instrumentation into the picometre spatial-precision regime [31,33,39].

Of these techniques, the most applicable in the context of 4D-STEM are the ones which deal with higher-dimensional data [35,40,41], and those which can compensate for fully non-linear scanning corrections (not just affine). To extend these techniques we first take a moment to consider the data-dimensionality, as illustrated in Figure 2.

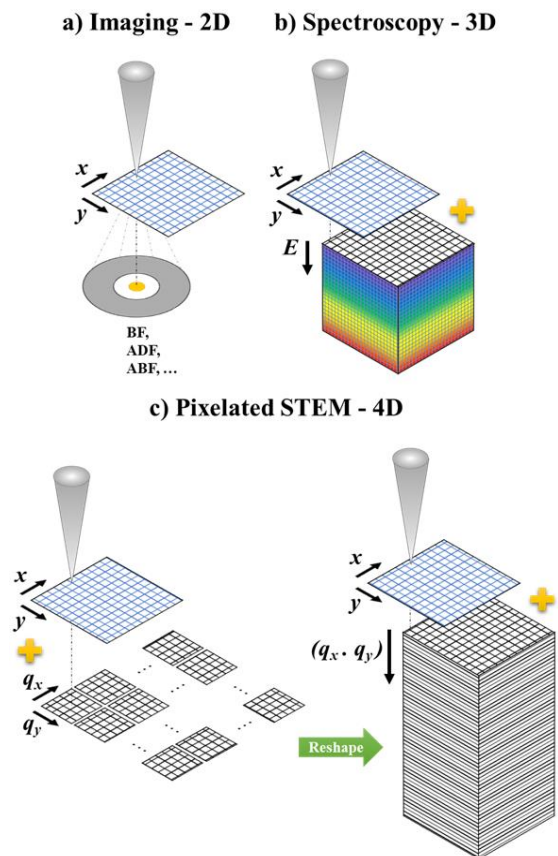

*Figure 2. Schematic of the dimensionality of various STEM data types. Conventional integrating-type imaging detectors such as BF, ADF, or ABF, (a) yield two-dimensional (2D) data. Spectroscopic techniques such as EELS or EDX add an energy dimension to this (b). Spatially resolved diffraction or ptychographic recordings with pixelated sensors record four-dimensional data (c). For all these techniques, multiple scan-frames may be recorded to form a series, in which case the dimensionality of the data is increased by 1.*

For conventional imaging (BF, ADF etc.), a two-dimensional array of values is recorded as a function of probe-position. There may be two or more of these detectors in use simultaneously, but each of these represents only one 2D array. Spectrum-imaging increases this data dimensionality to the third dimension by adding some energy axis. In this case, an additional spectrum dimension is recorded at every probe position in the 2D raster scan. The dimensionality of this is then the same as the case of energy filtered TEM imaging (EFTEM) but here we concentrate on the scanned case. In such 'spectrum volumes', every energy 'slice' in our scanned data volume has the same lateral real-space distortions and this is key to their eventual correction [40,41].

4D-STEM, as the name suggests, takes this one dimension further, with every real-space probe position corresponding to a two-dimensional image in the camera plane. If the time dimension of the series is also counted, this 4D-series may be considered as a single overall 5D dataset. Analogous to registering a series of spectrum-images, the 5D dataset will be registered in the two spatial dimensions and then projected along the time dimension into an improved 4D-STEM dataset. The complication is now that the detector signal is not a scalar, but a 2D array of pixel values in the momentum plane of the detector.

For each pixel in this camera plane, TEM-STEM reciprocity allows us to visualize this as a series of tilted illumination images all recorded simultaneously [42], where each tilted image possesses an identical real-space field-of-view and real-space scan-distortions. Additionally, this same reciprocity argument allows us to reshape the 4D array down to only three dimensions, where the third dimension now represents just a pixel-indexing number corresponding to that series of tilted images as each tilt can be processed as an independent data set (Figure 2).

Once the data are reshaped, they can be distortion-corrected using the existing approach followed by the SmartAlign algorithm [33,41]. This approach differs from some others in that it is not limited to affine only correction [31], does not require atomic-column peaks for distortion analysis [35], and does not require all the data to be opened in RAM simultaneously during the distortion correction [40]. Furthermore, the scan-diagnosis is compatible using data both with or without 90° scan-rotation increments [31,34].

After correction of this series of 3D data sets, they are reshaped back into a series of 4D data items and can be projected along (summed over) the series dimension to obtain an improved 4D-STEM data item or looked at individually as a registered time-series of 4D-STEM. In the following we will concentrate on the former option to show the improved characteristics of the projected 4D map, but the subsequent processing of the data after registration is the same in all cases and equal to a conventional 4D-STEM data set.

After correction, the data is reshaped back from the 3D form to a 4D form for onward processing with the researcher's choice of phase- or field-reconstruction algorithm [43–45].

## Methods

In this work, multi-frame 4D-data acquisition and registration is demonstrated for two 4D-STEM imaging techniques: 1) center-of-mass (COM) imaging and 2) non-iterative electron ptychography.

The COM E-field data was recorded using a Nion HERMES microscope located at Humboldt-Universität zu Berlin. The microscope was operated at 60 kV for each of the three examples.

The gadolinium aluminium gallium garnet (GAGG) data shown in Figure 3 and Figure 5 was acquired with a convergence angle of 36 mrad and an ADF inner-angle of 60 mrad, 128x128 detector pixels (windowed region of a 2048x2048 fiber-coupled Hamamatsu ORCA SCMOS camera) running at 1000 frames per second with an angular resolution of 1.2 mrad/pix, 110x126 scan points (128x128 before registration) with a sampling of 0.313 Å and a beam current of 43 pA.

In the case of the $WS_2$ sample shown in Fig. 4, the data was obtained using a convergence angle of 35 mrad with the same ADF inner-angle, 4D detector angular resolution, and windowing as before but running at 1600 frames per second, 68x96 scan points (128x128 before registration) with the same sampling but a current of 28 pA.

For the twisted bilayer graphene sample, the data was acquired with a convergence angle of 40 mrad and an ADF inner-angle of 50 mrad, a windowed 256x256 pixel detector region (binned by 4 to 64x64 pixels) of the prototype Dectris ELA direct detector (1030x514 pixels) [46] running at 4000 frames per second with an angular resolution of 2.5 mrad/pix, 219x189 scan points (256x256 before registration) with a sampling of 0.156 Å and a beam current of 85 pA.

The ptychographic data was recorded using a JEOL ARM200CF at the David Cockayne Centre for Electron Microscopy. A JEOL 4DCanvas fast pixelated detector (264x264 pixels) was used, which was operated at 4000 frames per second with four-fold binning (66x264 pixels) [47]. The microscope was operated at 80kV with a 31 mrad probe semi-convergence angle and 16pA beam current resulting in a dose of approximately $3.5 \times 10^5$ eÅ$^{-2}$ per scan-frame. A series of 21 frames were recorded of both pixelated STEM data and hardware-ADF images, resulting in a cumulative dose of $7.35 \times 10^6$ eÅ$^{-2}$. The real-space dimensions are 246x225 scan points (256x256 before registration) with a probe step size of 0.264 Å. The ptychographic phase reconstructions were obtained using the single-sideband method [14].

Using the physical-ADF signal for the registration and distortion diagnosis is convenient so long as there is adequate dark-field signal in each image. However, for the $WS_2$, GAGG, twisted bilayer, and ptychography data sets, these diagnoses were all performed using the virtual-ADF series synthesized from the 5D data. This alternative method is advantageous for two reasons.

Firstly, the virtual detector geometry can be optimized after acquisition to maximize the signal available for the image registration process. If the sample signal is weak at high scattering angles, a bright-field or DPC detector geometry can be assumed to generate a raw image series which can be used as input to the registration algorithm. Secondly, any synchronization issues between separate detectors can be avoided.

The drift and distortion diagnosis were performed using the SmartAlign algorithm [33], where importantly, the x-y sample shifts and the scan-distortion vector-fields are stored after the calculation is complete. The SmartAlign algorithm is described in detail elsewhere, but briefly, this approach involves comparing the intensity difference between the moving image and the reference image. The difference in the local gradients of the images then directs the pixel-wise offsets needed. This diagnosis is performed in an iterative manner until the full distortions are determined [33].

Some experimental optimizations may be made to maximize the quality of the eventual registration; for example, a small scan-rotation may be added during acquisition to rotate the low-order planes of the material away from the fast-scan direction. This avoids blank rows with no contrast and allows for more reliable diagnosis and compensation of scan-distortion. Additionally, at the camera-lengths required for the 4D cameras, the physical ADF detector may have a large inner-angle, causing the data to be somewhat noisy. In the case of noisy data due to large ADF inner-angles, gentle bandpass filtering was tried and did not seem to limit the precision of probe-offset vectors and therefore seems to be a viable option to enhance the registration.

In principle it is possible to correct a single 4D volume for scan-distortion if another reliable reference image is available (such as a fast multi-frame average ADF frame) [10,48], however by recording these data as an entire series the corrections can be made with no external prior knowledge applied.

# Results & Discussion

In this section, we first discuss the process of scan-distortion diagnosis, alignment and subsequent dose fusion and present the obtained new 'raw' 4D-STEM data set before discussing the improved quality of results obtained from this data by applying existing COM and ptychography workflows to it.

## Alignment & Scan-Distortion Diagnosis

The ADF frames acquired simultaneously to the 4D-STEM data were used for the scan-distortion diagnosis. This process yields a vector field with the same dimensions as each frame, such that the diagnosis provides the x-y vector needed to correct the probe-offset for every pixel over the field-of-view and in every frame. An example of these x-y vector data from one frame of the experimental series is shown in the Supplementary Information (Fig S1). For an ADF image, a scan-distortion on the scale of a single pixel would be easily visible in a strain analysis but would otherwise be considered qualitatively minor.

As single side-band ptychographic reconstruction relies on Fourier transforms (taken with respect to real-space) this spatially-distributed distortion becomes a phase corruption in Fourier-space and may significantly affect the precision of both aberration diagnosis and reconstruction [21].

For COM analysis of the data, these distortions are a major problem as the E-field vectors critically rely on the precise position of the probe. This can be rationalized by the fact that the E-field (whose deflecting influence on the probe is directly measured) diverges towards the atom centers and then flips sign on crossing it, which is a distinctly different behavior from other signals that are typically collected. The finer the probe, the more important is a precise knowledge of the (actual) probe position as the (divergent) field is not smeared out so much and leads to a more abrupt change of COM vector on crossing the atom center. Therefore, COM greatly benefits from the registration of 4D series.

## Dose-Fusion after Alignment

After rigid translation in real-space, the data are cropped to their common field-of-view and then aligned using non-rigid registration. The 4D datasets corrected with the same diagnoses also retain this same field-of-view.

Figure 3 shows example Ronchigrams from both a single 4D dataset and a dose-fused dataset (registered and integrated) of a gadolinium aluminium gallium garnet (GAGG) sample. By comparing Ronchigrams from a single 4D map (a) and ten fused maps (b) of a column position (blue) and an off-column position (orange) it can be seen that the dose-fused data has an improved noise level without exhibiting any artefacts from the registration process. Fine details in the Ronchigrams are clearly conserved and the only visible influence of the process is the augmentation of the number of electrons and subsequent reduction of Poisson noise. The enhancement of the data through (sub-pixel) correction of the actual probe position is not fully revealed from this figure but will become evident in the subsequent analyses.

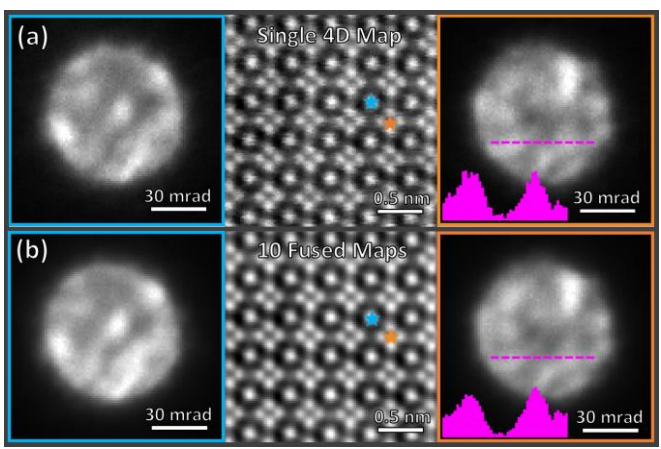

*Figure 3. Individual Ronchigrams extracted from a column position (blue) and a position in between columns (orange) of a single 4D map (a) and the sum of a non-rigid registration (b) from a gadolinium aluminium gallium garnet (GAGG) sample. The positions are indicated with stars in the corresponding HAADF images in the center. The pink dashed line indicates the position of the profiles that are also plotted in pink and illustrate the reduction of noise in the fused data compared to a single map. The data set is the same as for Figure 5.*

## COM Analysis

Here we present three illustrative examples to show the applicability of the approach; a monolayer material ($WS_2$) and a bulk-like crystal (GAGG) each acquired using an optically coupled CMOS detector, and a twisted bilayer graphene data-set acquired using the Dectris ELA prototype direct detector [46]. Detailed parameters of the data sets can be found in the Methods section.

Figure 4 depicts a virtual ADF image, the color-coded E-field map obtained from the COM analysis and the projected charge density (calculated from the E-field) of a single 4D-STEM data set (a) and of six registered sets (b) of a $WS_2$ monolayer sample. The difference between (a) and (b) is striking, but while in the case of the ADF image the two defects are quite easy to spot even for the single map, the difference between top and bottom is even greater for the E-field and especially the charge density maps. In the case of the E-field maps, having a vector field that changes strongly (atoms are rotationally symmetric in ADF but not in the E-field map) enhances the detrimental effect of the distortions. For the charge density maps, the values depend on the local surroundings and therefore distortions in the spatial coordinate even lead to differences in local magnitudes. The dose fusion leads to high-quality data from monolayer materials even when using an optically coupled detector.

In Figure 5 an analogous analysis for the GAGG sample is shown, whose Ronchigrams were already qualitatively compared in Figure 3. The results from a single map (a) and ten fused maps (b) of this bulk-like sample (also acquired with an optically coupled camera) are shown in comparison. The stronger scattering signal and improved stability from this thicker sample compared to the previous one leads to less artefacts in the maps in (a), but (b) still shows higher fidelity of the positions and, more importantly here, a strongly enhanced SNR that clearly brings out smaller features of the structure.

Figure 6 shows data sets from a twisted bilayer graphene sample, this time acquired using a direct electron detector (Dectris ELA prototype [46]). Due to the relatively fast acquisition speed (4000 images per second in this example), the distortions in the maps obtained from the single data set, shown in Figure 6(a), seem not so strong, but the signal of this weakly scattering sample is obviously very low. Conversely, the fusion of six maps, as shown in Figure 6(b), leads to a result that allows for a clear observation of the E-field distribution and thus also of the charge density.

In summary, the dose-fusion allows one to obtain high quality 4D-STEM data sets even for slow and noisy optically-coupled detectors and, in the case of novel fast direct detectors, allows to harness their full speed and acquire multiple maps instead of having to increase the dwell time of a single map to boost the signal and thus introduce more artefacts from instabilities.

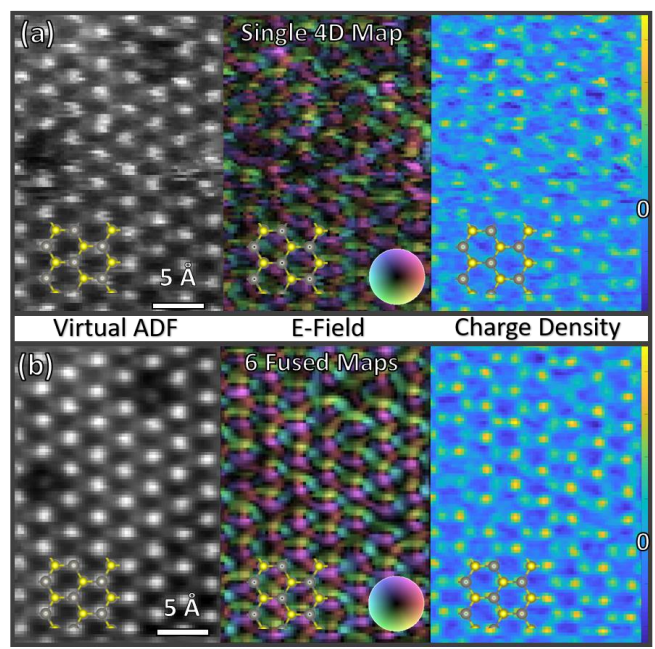

*Figure 4. Comparison of single 4D map (a) and 6 fused (non-rigidly registered) data sets (b) of a $WS_2$ monolayer with two defects: virtual ADF image (left), E-field map (center), and charge-density distribution (right).*

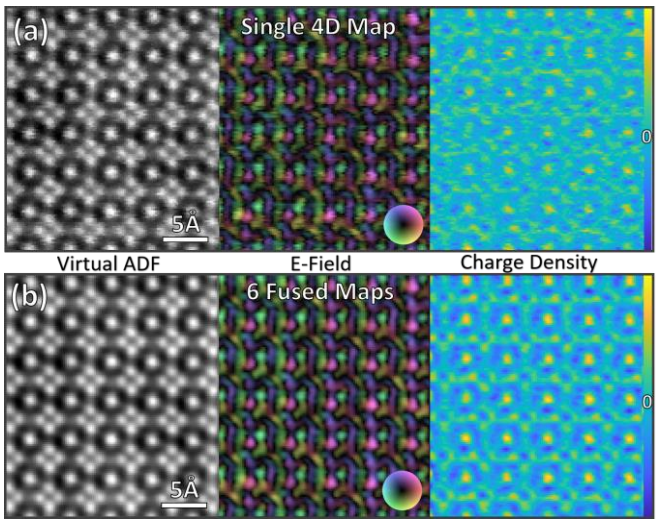

*Figure 5. Single 4D data set (a) and sum of 10 (non-rigidly) registered maps (b) of a gadolinium aluminium gallium garnet (GAGG). Depicted are virtual ADF images (left), color-coded E-Field map (center) and projected charge density (right).*

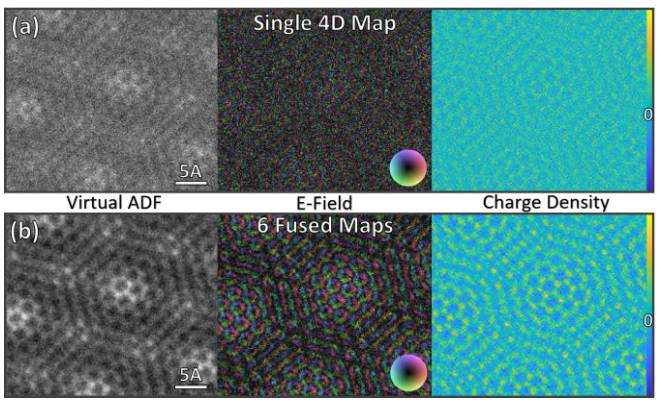

*Figure 6. Comparison of single 4D-STEM data set (a) and the sum of 6 fused (non-rigidly registered) maps (b) of a twisted bilayer graphene sample: virtual ADF image (left), E-field map (center), and charge-density distribution (right).*

Increasing the number of frames of 4D data contributing to the sum increases the effective total electron dose used in each type of reconstruction, thus an increase in precision is expected.

The same is true of the electric-field map precision, where the curl operator of the field-map can be used to indicate the fidelity of the data. In this example the curl was reduced from 0.0390 to 0.0173, an improvement of 2.25x. The COM analysis greatly benefits from correcting the probe positions due to the peculiar properties of the E-field signal on crossing atom positions.

## Ptychography Analysis

To demonstrate the versatility of non-rigid registration of 4D-STEM data, we also compared the performance of non-iterative, single side-band electron ptychography reconstructions for individual and dose-summed registered data sets. The ADF images and phase reconstructions for both the single frame and dose-summed data are shown in Figure 7. Monolayer graphene is a weak-phase-object and thus provides poor dark-field signal in the detector plane as seen from the ADF in Figure 7(a). A combination of registration and dose-summing is required to reveal the atomic columns with high spatial precision (Figure 7(b)). The application of electron ptychography to single data sets as shown in Figure 7(c) can reconstruct phase maps with a much higher SNR than the simultaneously acquired ADF images. However, the fidelity of the reconstruction is hampered by the noticeable phase variations between identical atoms. By applying multi-frame acquisition, non-rigid registration methods and dose-summing before running the electron ptychography reconstruction, as shown in Figure 7(d), the SNR, and hence the spatial precision of the phase reconstructions, can be improved significantly. The spatial frequency power spectra for these 4D data sets are shown in Supplementary Figure S2.

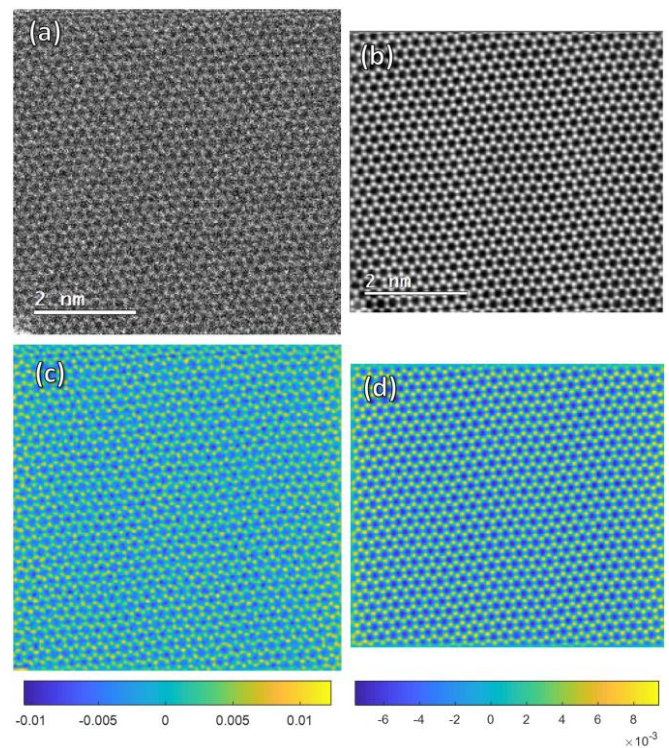

*Figure 7. Multi-frame ptychography of a graphene monolayer. (a) Typical single ADF frame (10 of 21) from the raw ADF series. (b) Aligned and non-rigid registered average ADF image cropped to 246 x 225 px. (c) Single side-band (SSB) phase image from an individual frame. (d) SSB phase image from the aligned and accumulated data set. Note: after alignment, cropping the common area present in all frames results in a small loss of image field-of-view.*

To quantitatively demonstrate the improvement to SNR and phase precision for the ptychographic data, an integrated squared-phase cross-section (ISPCS) was evaluated for each atom in the phase reconstructions. The ISPCS is calculated by integrating the squared-phase values for a Voronoi cell around each atom. The reason that the squared-phase was chosen instead of the phase is as follows. The positive phase contributions from the atomic potentials are completely cancelled out by the negative phase contributions from the surrounding vacuum. As such, the integrated phase over a Voronoi cell should equal zero. By squaring the phase, the signal from both the atomic potentials and surrounding vacuum is positive, resulting in a non-zero ISPCS value in each Voronoi cell. Figure 8 shows histograms of the ISPCS from a single 4D-frame (orange) and the dose-fused non-rigid registered ensemble (green).

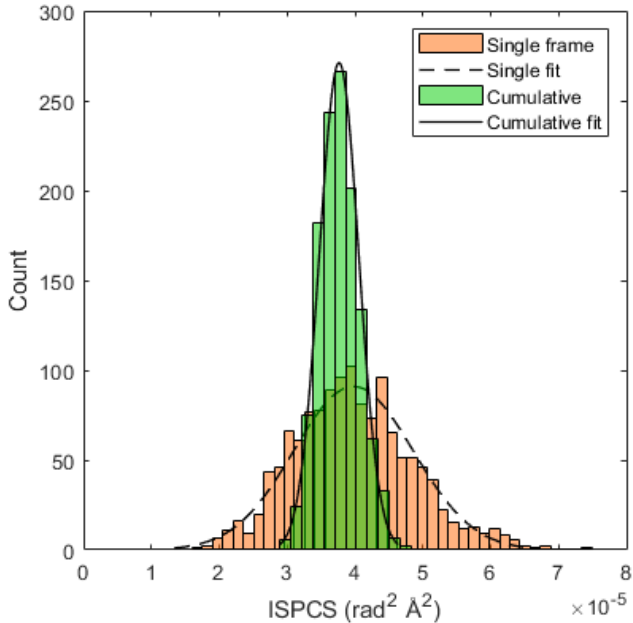

*Figure 8. Histograms of the integrated squared-phase cross-section (ISPCS) for the central frame of the series (frame 10 of 21, orange) and for the non-rigid registered and summed data (green). A normal distribution has been fitted to each histogram.*

For a single frame, the ISPCS values for all the carbon atoms in the image have a standard deviation of $8.80 \times 10^{-6}$ rad$^2$ Å$^2$. After dose fusion of 21 frames, the standard deviation for the same set of carbon atoms fell to $2.91 \times 10^{-6}$ rad$^2$ Å$^2$. The standard deviation of the ISPCS was reduced by more than a factor of 3 with multi-frame acquisition and registration. This provides a significant improvement in feature interpretability as shown in Figure 7, which is particularly important for distinguishing subtle atomic number differences and bonding effects in light-element samples. It should be noted that the improvement in phase-precision is less than the factor of $\sqrt{21}$ that would be expected from Poisson statistics for incoherent imaging methods. Ptychographic reconstructions do not follow Poisson statistics, and this might be the explanation for the slightly lower precision improvement. Although the mode of the two ISPCS values are similar, the mean can be seen to differ slightly. The single frame distribution is also somewhat asymmetric, which may be associated with the way scanning distortions affect the ptychographic reconstruction.

In addition to improving the phase-precision, the increased SNR of registered and dose-summed 4D-STEM data can improve the contrast of high-resolution information in the phase reconstructions. This can be seen in the Fourier transform of the real-space phase maps as shown in Supplementary Figure S5. For the single 4D-STEM data set, the 3$^{rd}$ order spots are the highest spatial-frequency observable indicating a resolution of 1.05Å. After the distortion-correction and signal-fusion, the 4$^{th}$ order spots at 0.79Å are clearly visible and the 5$^{th}$ order spots at 0.71Å are marginally visible. This represents a resolution improvement of approximately 48%.

The 4D multi-frame alignment not only increases the dose available to the final reconstruction, but also increases the dose presented at the stage of the post-processing aberration diagnosis and correction procedures which can be performed as part of the ptychographic workflow [21]. Supplementary Figures S3 and S4 show example disk-overlap plots for a single-scan and the 21-frame dose-fusion volumes.

The first-order disk-overlaps (Supplementary Figures S3 and S4, top row) corresponding to ≈4.8nm$^{-1}$ (or 0.65α) have a SNR that is ample for aberration diagnosis in both the single-frame and dose-fused data. The third-order overlaps however (Supplementary Figures S3 and S4, bottom row), which correspond to ≈9.5nm$^{-1}$ (or 1.28α), show very little information in the single-frame but a significant improvement in SNR is seen after dose-accumulation and scan-distortion correction. After dose-fusion, the chromatic damping-envelope is even visible in the amplitude of the third-order overlaps (Supplementary Figure S3, bottom-right), which can be utilized to reduce the chromatic defocus spread of ptychographic reconstructions [14,49]. The improved SNR in the phase of the overlaps (Supplementary Figure S4), especially at higher spatial frequencies, improves the performance of the aberration diagnosis and correction procedures which can be used along with ptychography to provide aberration-free phase reconstructions.

For both the COM and ptychography examples investigated, the improvements in electric-field self-consistency, phase-precision, or spatial-resolution, all depend on the precise scan-position errors and their interplay with the nuclear coordinates. A more expansive study is needed to determine the mathematical framework of what improvements should be expected with increasing numbers of frames.

## Conclusions

In summary, it was demonstrated that the data fusion of non-rigidly registered 4D-STEM series enhances the signal-to-noise ratio (SNR) and improves the fidelity of scan positions without introducing artefacts. This enhanced 'raw' data can then be processed like any other conventional 4D-STEM data set but with drastically enhanced results.

It was shown that the improved scan positions avoid phase corruption in the case of ptychography and greatly enhance the result of COM analysis, as the calculated E-fields strongly rely on the beam positions which are significantly imperfect in experimental 4D-STEM data sets. We observed a factor of 3 improvement in squared-phase precision, a 48% improvement in spatial resolution as well as improved robustness against cold-FEG emission instability.

As 4D-STEM data is severely noise-limited and this framework offers a way on how to extend the acquisition time and thus the dose and SNR without running into the typical problems associated with sequential data acquisition, it should be a way on how to reveal signals that are typically too weak to observe well, like the real space observation of bonding influences. To facilitate this, it is ideal to combine this post-processing of 4D-STEM series with a simple (rigid) drift correction during acquisition, as not to lose the field of view during acquisition.

Although drift correction using reference frames has long been possible [48], we have instead demonstrated an internal-reference approach based on series acquisition [33].

While ptychographic and center-of-mass type datasets have been the focus of this manuscript all the methods described are nevertheless applicable to other multi-frame 4D measurements such as STEM diffractive imaging, STEM symmetry imaging [50], or magnetic field imaging [12,51]. Our approach is not limited to high-resolution data; this was chosen as an especially challenging example, but other 4D-STEM scenarios like scanning nano-beam maps can be treated in the same way.

## Acknowledgments

The authors acknowledge use of characterization facilities within the David Cockayne Centre for Electron Microscopy, Department of Materials, University of Oxford and in particular the EPSRC (EP/K040375/1 "South of England Analytical Electron Microscope") and additional instrument provision from the Henry Royce Institute (Grant reference EP/R010145/1). LJ acknowledges SFI grants URF/RI/191637 and AMBER2-12/RC/2278_P2. BH and CTK gratefully acknowledge the German science foundation (DFG) funding in the form of the 'BerlinEM Network' project (KO 2911/12-1) and the SFB951 HIOS (Projektnummer 182087777). The authors thank Alisa Ukhanova (Lomonosov Moskau State University) and Oleg Busanov (Fomos-Materials) for providing and preparing the GAGG sample, Xiaomin Xu & Norbert Koch (Humboldt-Universität zu Berlin) for the $WS_2$ sample and Andreas Mittelberger (Nion Co.) for providing a 4D-STEM plugin for the Nion Swift software.

# Supplementary Information

## Scan-distortion Diagnosis

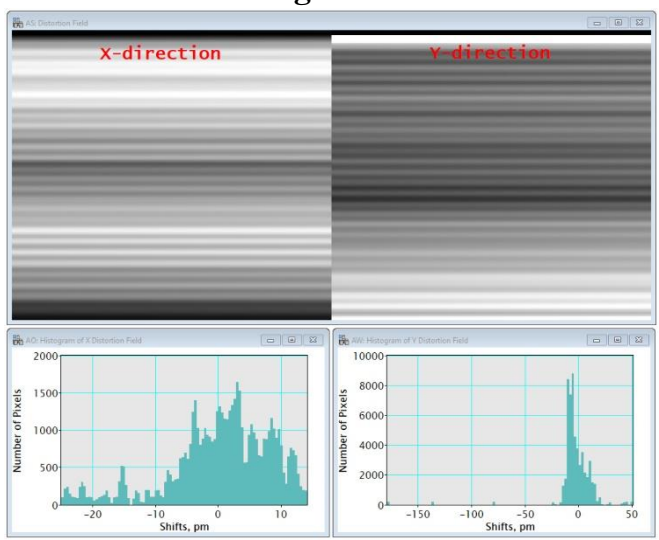

*Figure S1. Scan-distortion analysis of an example ADF from the image-series. The x and y direction components of the scan-distortion vector-field are shown, as well as histograms of their respective pixel values. The x-direction distortions are largely within ± 10 pm. The y-direction is similar except for a vertical compression visible at the top of the raw image in Figure 7.*

In Fig S1 the probe-offsets are represented as the shifts in x and y needed to correct the probe-positions. A similar vector-field exists for each of the 21 scan-frames in the dataset and these are stored with each image as meta-data. From these plots, we can determine that the approximate magnitude of the random probe-offsets is around ±10 pm in x and ±20 pm in y. With a pixel-size of 0.264 Å in the raw data, these random shifts are on the same order-of-magnitude.

## Spatial-frequency Power Spectrum

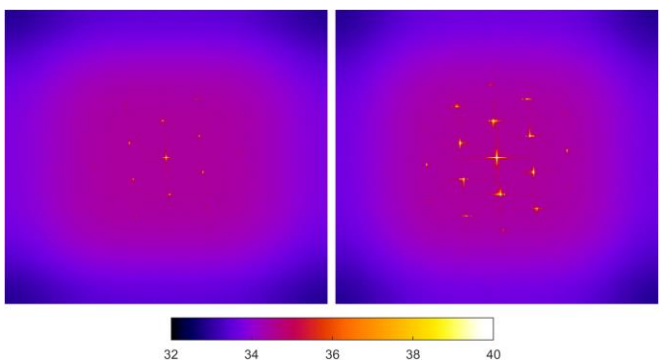

*Figure S2. Power-spectrum across the spatial dimensions of the 4D data-volumes; (left) from a single scan-frame, (right) from the cumulative 21 frames.*

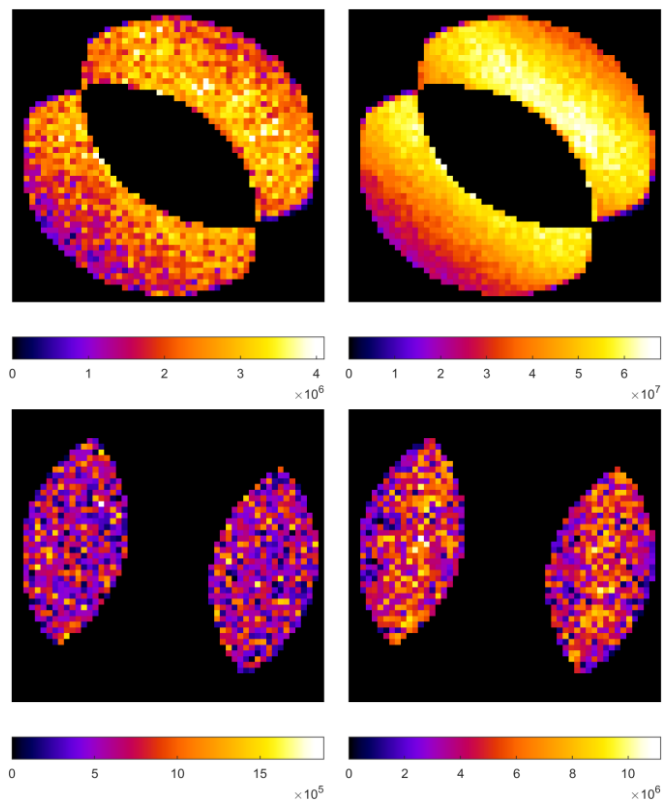

*Figure S3. Amplitude plots of the first and third order disk-overlaps corresponding to Figure 7. After dose-fusion, the chromatic damping envelope [14,49] is visible even for the third order overlaps.*

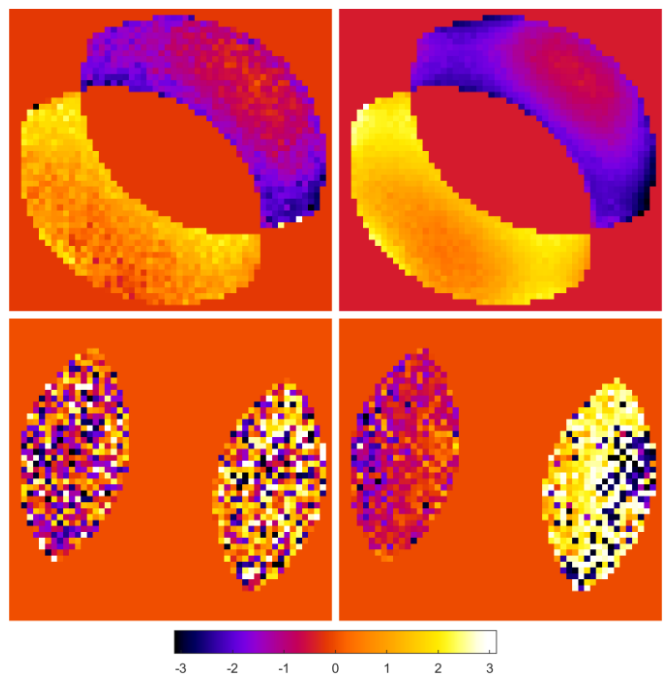

*Figure S4. Phase plots of the disk overlap regions for one example of the first-order Fourier components of the graphene lattice (top) and for a third-order component (bottom). For each one, both the single scan-frame (left) and the cumulative equivalent (right) are shown. For fair comparison, aberration correction and phase unwrapping has been applied to each. Amplitude plots for these overlaps are shown in the supplementary information.*

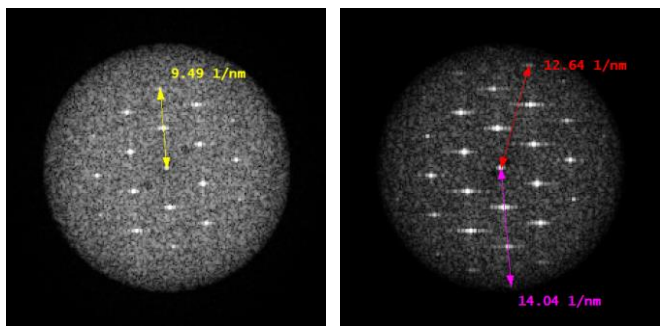

*Figure S5. Fourier transforms of the real-space phase-maps from (left) and single 4D-scan, and (right) from the fusion of 21 scans.*